\documentclass[twocolumn,amsmath,amssymb]{revtex4}
 
\usepackage{graphicx}
\usepackage{dcolumn}
\usepackage{bm}

\begin{document}
\newcommand{\be}{\begin{equation}}
\newcommand{\ee}{\end{equation}}
\newcommand{\bea}{\begin{eqnarray}}
\newcommand{\eea}{\end{eqnarray}}
\newcommand{\n}{\nonumber\\}

\title{\bf Structural Properties of Planar Graphs of Urban Street Patterns}

\author{ Alessio Cardillo$^{1}$, 
         Salvatore Scellato$^{2}$,  
         Vito Latora$^{1}$ and 
         Sergio Porta$^{3}$}

\affiliation{$^{1}$ Dipartimento di Fisica e Astronomia, 
Universit\`a di Catania, and INFN Sezione di Catania, Italy}

\affiliation{$^{2}$ Scuola Superiore di Catania, Italy}

\affiliation{$^{3}$  Dipartimento di Progettazione dell'Architettura, 
Politecnico di Milano, Italy}

\date{\today}

\begin{abstract}
Recent theoretical and empirical studies have focused on the structural properties 
of complex relational networks in social, biological and technological systems. 
Here we study the basic properties of twenty 1-square-mile samples of street patterns 
of different world cities. 
Samples are represented by spatial (planar) graphs, i.e. valued graphs 
defined by metric rather than topologic distance and where street intersections are 
turned into nodes and streets into edges. 
We study the distribution of nodes in the 2-dimensional plane.  
We then evaluate the local properties of the graphs by measuring the 
meshedness coefficient and counting short cycles ( of three, four and five edges), 
and the global properties by measuring global efficiency and cost. 
As normalization graphs, we consider both minimal spanning trees (MST) and 
greedy triangulations (GT) induced by the same spatial distribution of 
nodes. The results indicate that most of the cities have evolved into 
networks as efficienct as GT, although their cost is closer to the one 
of a tree. 
An analysis based on relative efficiency and cost is 
able to characterize different classes of cities. 
\end{abstract}
\vspace{0.5cm}
\maketitle

\section{Introduction}
During the last decade, the growing availability of large databases,
the increasing computing powers, as well as the development of reliable 
data analysis tools, have constituted a better machinery 
to explore the topological properties of several complex networks  
from the real world ~\cite{bareport,newmanreview,vespignanibook,report}. 
This has allowed to study a large variety of systems as diverse as 
social, biological and technological. The main outcome of this  
activity has been to reveal that, despite the inherent
differences, most of the real networks are characterized by
the same topological properties, as for instance relatively small
characteristic path lengths and high clustering coefficients ( the 
so called {\em small-world property}) \cite{ws98}, 
{\em scale-free} degree distributions \cite{bascience99}, 
degree correlations \cite{pvv01}, and the presence of motifs \cite{alonScience02} 
and community structures \cite{wasserman94}. 
All such features make real networks radically different from
regular lattices and random graphs, the standard topologies usually 
used in modeling and computer simulations.  
This has led to a large attention towards the comprehension  
of the evolution mechanisms that have shaped the topology of a real network, 
and to the design of new models retaining the most significant 
properties observed empirically.

{\em Spatial networks} are a special class of complex networks 
whose nodes are embedded in a two or three-dimensional Euclidean space 
and whose edges do not define relations in an abstract space 
(such as in networks of acquaintances or collaborations between
individuals), but are real physical connections \cite{report}. 
Typical examples include neural networks \cite{sporns03}, 
information/communication networks \cite{yook02,lm05a}, 
electric power grids \cite{kcal05} and transportation systems 
ranging from river \cite{pitts}, to airport
\cite{amaralairport,barrat04a} and street
\cite{porta_epb2} networks. 
Most of the works in the literature, with a few relevant exceptions 
\cite{yook02,ants04,newman04},  
have focused on the characterization of the
topological properties of spatial networks, while 
the spatial aspect has received less attention, when not neglected 
at all. 
However, it is not surprising that the topology of such systems is 
strongly constrained by their spatial embedding. 
For instance, there is a cost to pay for the existence of long-range 
connections in a spatial network, this having important consequences on the 
possibility to observe a small-world behavior. 
Moreover, the number of edges that can be connected to a single 
node is often limited by the scarce availability of physical space, 
this imposing some constraints on the degree distributions.  
In few words, spatial networks are different from other complex networks 
and as such they need to be studied in a different way. 

In this paper we focus on a particular class of spatial networks:  
{\em networks of urban street patterns}. We consider a database of 
1-square mile samples of different world cities and for 
each city we construct a spatial 
graph by associating nodes to street intersections and edges to streets. 
In this way, each of the nodes of the graph is given a location 
in a 2-dimensional square, and a real number, representing the 
length of the corresponding street,   
is associated to each edge. By construction, the resulting 
graphs are {\em planar graphs}, i.e. graphs forming nodes whenever 
two edges cross. 
After a previous work on the distribution of centrality measures 
\cite{porta_centrality}, here we present a comparative study of the 
basic properties of spatial networks of different city street patterns. 
In particular we evaluate the characteristics of the graphs both at a global 
and at a local scale. 
The main problem with spatial graphs is that, in most of the 
cases, the random graph or the complete graph are no more a good 
way to normalize the results. In fact, the common procedure in 
relational (non-spatial) complex networks is to compare the properties 
of the original graph derived from the real system 
with those of some randomized versions of the graph, 
i.e. of graphs with the same number of nodes and links as the original 
one, but where the links are distributed at random. 
This is, for instance, the standard way proposed by Watts and 
Strogatz in Ref.~\cite{ws98} to assess whether a real 
system is a small world. 
One quantifies the structural properties of the original graph 
by computing its characteristic path length $L$ and  clustering 
coefficient $C$, where $L$ measures the typical separation
between two vertices in the graph (a global property), whereas $C$ 
measures the cliquishness of a typical neighbourhood (a local property). 
Then, the graph is a small-world if $L$ assume a value close 
to that obtained for the randomized version of the graph, $L_{rand}$, 
while the value of $C$ is much larger than $C_{rand}$. 
Similarly, in the efficiency-based formalism proposed in 
Refs.~\cite{lm01,lm03}, a small-world network is defined as 
a system being extremely efficient in exchanging information both at 
a global and at a local scale. 
Again, the values of global and local 
efficiency are compared with those obtained for a randomized 
version of the graph. 
A similar method is used in the counting of short cycles or specific 
motifs in a graph representing a real system \cite{alonScience02}. 
The research of the motifs and cycles is based on matching algorithms 
counting the total number of occurrences of each motif and each 
cycle in the original graph and in the randomized ones. 
Then, a motif or a cycle is statistically significant if it appears 
in the original graph at a number much higher than in the randomized 
versions of the graph. 
\noindent
In a planar graph, as those describing urban street patterns, 
the randomized version of the graph is not significative because 
it is almost surely a non-planar graph due to the edge crossings 
induced by the random rewiring of the edges.  
Moreover, because of the presence of long-range links, 
a random graph corresponds to an extremely costly street pattern  
configuration, where the cost is defined as the sum of street 
lengths \cite{lm03}. The alternative is to compare urban street patterns  
with grid-like structures. Following Ref.~\cite{ants04}, we shall consider 
both {\em minimum spanning trees} and {\em greedy triangulations} 
induced by the 
real distribution of nodes in the square. 
Spanning trees are the planar graphs with the minimum 
number of links in order to assure connectedness, while 
greedy triangulations are graphs with the maximum number 
of links compatible with the planarity. 
Spanning trees and greedy triangulations will serve as the two extreme 
cases to normalize the structural measures we are going to compute.

\section{Networks of urban street patterns} 

The database we have studied consists of twenty 1-square mile samples of different 
world cities, selected from the book by Allan Jacobs \cite{jacobs}. 
We have imported the twenty maps into a GIS (Geographic Information System) 
environment and constructed the correspondent spatial graphs 
of street networks by using a road-centerline-between-nodes format \cite{dalton}. 
Namely, each urban street pattern is trasformed into a undirected, 
valued (weighted) graph $G = ({\cal N},{\cal L})$, embedded in 
the 2-dimensional unit square. 
In Fig.~\ref{maps} we show the case for the city of Savannah: 
in the upper-left panel we report the original map, and in upper-right 
panel the obtained graph. 
$\cal N$ is the set of $N$ nodes representing 
street intersections and characterized by their positions 
$\{x_i,y_i\}_{i=1,...,N}$ in the square. ${\cal L}$ is the set 
of  $K$ links representing streets. The links follow 
the footprints of real streets and are associated a set of real positive 
numbers representing the street lengths, $\{l_k\}_{k=1,...,K}$. 
The graph is then described by the adjacency $N \times N$ 
matrix $A$, whose entry $a_{ij}$ is equal to 1 when there is an edge 
between $i$ and $j$ and 0 otherwise, and by a $N \times N$ matrix $L$,
whose entry $l_{ij}$ is equal to the 
length of the street connecting node $i$ and node $j$.
In this way both the topology and the geography (metric distances) 
of the system will be taken into account.
A list of the considered cities is reported in table \ref{table1}, 
together with 
the basic properties of the derived graphs. 
The considered cases exhibit striking differences in
terms of cultural, social, economic, religious and geographic context. 
In particular, they can be roughly divided into two large classes: 
1) patterns grown throughout a largely self-organized, fine-grained historical process, 
out of the control of any central agency; 
2) patterns realized over a short period of time as the result of a single 
plan, and usually exhibiting a regular grid-like, structure. 
Ahmedabad, Cairo and Venice are the most representative examples 
of self-organized patterns, while Los Angeles, Richmond, and San Francisco 
are typical examples of mostly-planned patterns.
We have selected two different parts of the city of Irvine, CA, 
(named Irvine 1 and Irvine 2) for two highly diverse kinds of urban fabrics: 
the first is a sample of an industrial area showing enormous blocks 
with few intersections while the second is a typical residential 
early sixties ``lollipop'' low density suburb based on 
a tree-like layout with a lot of dead-end streets.
The differences between cities are already evident from the basic 
properties of the derived graphs. 
In fact, the number of nodes $N$, the number of links $K$, and the cost of the 
wiring, defined as the sum of street lenghts:
\begin{equation} 
 Cost = \sum_{i,j} a_{ij} l_{ij}
\label{eq:cost}
\end{equation}
and measured in meters,  
assume widely different values, notwithstanding the fact we have  
considered the same amount of land. Notice that 
Ahmedabad has 2870 street intersections and 4387 streets in a surface of 1-square mile, 
while Irvine has only 32 intersections and 37 streets. 
Ahmedabad and Cairo are the cities with the largest cost, while the cost is 
very small (less than 40000 meters) in Barcelona, Brasilia, Irvine, 
Los Angeles, New Delhi, New York, San Francisco, Washington and Walnut Creek. 
A large difference is also present in the average edge (street) length 
$\langle l \rangle $, 
that assumes the smallest values in cities as Ahmedabad, 
Cairo and Venice, and the largest value in San Francisco, 
Brasilia, Walnut Creek and Los Angeles. 
In Ref.~\cite{porta_centrality} we have studied the edges length 
distribution $P(l)$ for the two different classes of cities, showing  
that self-organized cities  show single peak distributions, 
while mostly planned cities exhibit a multimodal distribution, due to
their grid pattern. 
%
\begin{table}
\begin{tabular}{l|l|c|c|c|c|c|}
  & CITY             &   N  &  K    &  Cost  &  $\langle l \rangle$ & $D_{box}$  \\
\tableline
1    & Ahmedabad     & 2870 & 4387  & 121037 &  27.59 & 1.92   \\
2    & Barcelona     & 210  & 323   &  36179 & 112.01 & 1.99   \\ 
3    & Bologna       & 541  & 773   &  51219 &  66.26 & 1.95   \\
4    & Brasilia      & 179  & 230   &  30910 & 134.39 & 1.83   \\
5    & Cairo         & 1496 & 2255  &  84395 &  37.47 & 1.82   \\
6    & Irvine 1      & 32   & 36    &  11234 & 312.07 &  --    \\
7    & Irvine 2      & 217  & 227   &  28473 & 128.26 & 1.81    \\
8    & Los Angeles   & 240  & 340   &  38716 & 113.87 & 1.90    \\
9    & London        & 488  & 730   &  52800 &  72.33 & 1.94    \\
10   & New Delhi     & 252  & 334   &  32281 &  96.56 & 1.85    \\
11   & New York      & 248  & 419   &  36172 &  86.33 & 1.72   \\
12   & Paris         & 335  & 494   &  44109 &  89.29 & 1.88   \\
13   & Richmond      & 697  & 1086  &  62608 &  57.65 & 1.78    \\
14   & Savannah      & 584  & 958   &  62050 &  64.77 & 1.85   \\
15   & Seoul         & 869  & 1307  &  68121 &  52.12 & 1.87    \\
16   & San Francisco & 169  & 271   &  38187 & 140.91 & 1.90   \\
17   & Venice        & 1840 & 2407  &  75219 &  31.25 & 1.81   \\
18   & Vienna        & 467  & 692   &  49935 &  72.16 & 1.88  \\
19   & Washington    & 192  & 303   &  36342 & 119.94 & 1.93   \\
20   & Walnut Creek  & 169  & 197   &  25131 & 127.57 & 1.80   \\
\end{tabular}
\caption{Basic properties of the planar graphs obtained from the twenty 
city samples considered. $N$ is the number of nodes, $K$ is the number of 
edges, $Cost$ and $\langle l \rangle $ are respectively the total length 
of edges and the average edge length (both expressed in meters), 
$D_{box}$ is the box-counting fractal dimension.} 
\label{table1} 
\end{table}
We now have gone deeper into the characterization of the distributions of 
nodes (street intersections) in the unit square: we have calculated 
the fractal dimension of the distributions, by using the box counting method 
\cite{strogatz}. 
In all the samples, except Irvine 1 that is too small to draw any conclusion,  
we have found that the nodes are distributed on a 
fractal support with a fractal dimension ranging from 1.7 to 2.0. 
This result is similar to that obtained by Yook et al. 
for the spatial distribution of the nodes of the Internet, considered 
both at the level of routers and at the level of autonomous systems ~\cite{yook02}.

\subsection{Minimum Spanning Tree (MST) and Greedy Triangulation (GT)}
\label{section_mstgt}

Planar graphs are those graphs forming vertices whenever two edges cross, 
whereas non-planar graphs can have edge crossings that do not 
form vertices \cite{west95}. 
The graphs representing urban street patterns are, by construction,  
planar, and we will then compare their structural properties with 
those of minimally connected and maximally connected planar graphs. 
In particular, following Buhl et al.~\cite{ants04}, we consider  
the Minimum Spanning Tree (MST) and the Greedy Triangulation (GT) 
induced by the distribution of nodes (representing street 
intersections) in the square. 
The {\em Minimum Spanning Tree (MST)} is the shortest tree which connects 
every nodes into a single connected component. By definition the 
MST is an acyclic graph that contains $K_{min}=N-1$ links. This is the minimum 
number of links in order to have all the nodes belonging to a single connected 
component \cite{west95}. 
At the other extreme, the maximum number of links, $K_{max}$, that can be accomodated  
in a planar graph with $N$ nodes (without breaking the planarity) 
is equal to $K_{max}=3N -6$ \cite{b98}. The natural reference graph should be then 
the {\em Minimum Weight Triangulation (MWT)}, which is the planar graph 
with the highest number of edges $K_{max}$, and that minimize   
the total length.  
Since no polynomial time algorithm is known to compute the MWT,  
we thus consider the {\em Greedy Triangulation (GT)}, that is based on 
connecting couples of nodes in ascending order of their distance 
provided that no edge crossing is introduced \cite{dikerson_gt}.  
The GT is easily computable and leads to a maximal connected planar 
graph, while minimizing as far as possible the total length 
of edges considered. 

To construct both the MST and the GT induced by the spatial distribution 
of points (nodes) $\{x_i,y_i\}_{i=1,...,N}$ in the unit square, we have 
first sorted out all the couples of nodes, representing all the possible 
edges of a complete graph, by ascending order of their length. 
Notice that the length of the edge connecting node $i$ and node $j$ 
is here taken to be equal to the Euclidean distance 
$d^{Eucl}_{ij} = \sqrt{ (x_i - x_j)^2 + (y_i - y_j)^2 }$.  
Then, to compute the MST we have used the Kruskal algorithm 
\cite{cormen,kruskal}. 
The algorithm consists in browsing the ordered list, 
starting from the shortest edge and progressing toward the longer ones. 
Each edge of the list is added if and only if the graph obtained after 
the edge insertion is still a forest or it is a tree. 
A forest is a disconnected graph in which any two elements are 
connected by at most one path, i.e., a disconnected ensemble of trees. 
(In practice, one checks whether the two end-nodes of the edge 
are belonging or not to the same component). 
With this procedure, the graph obtained after all the links of the ordered 
list are considered is the MST. In fact, when the last link is included in 
the graph, the forest reduces to a single tree. 
Since in the Kruskal algorithm an edge producing a crossing would also 
produce a cycle, following this procedure prevents for creating 
edge crossings.   
To compute the GT we have constructed a brute force algorithm 
based on some particular characteristics of planar GT \cite{dikerson_gt}. 
The algorithm consists in browsing the ordered list of edges 
in ascending order of length, and checking for each edge whether 
adding it produces any intersections with any other edge already 
added.

\begin{figure}[htb]
\includegraphics[width=8.cm]{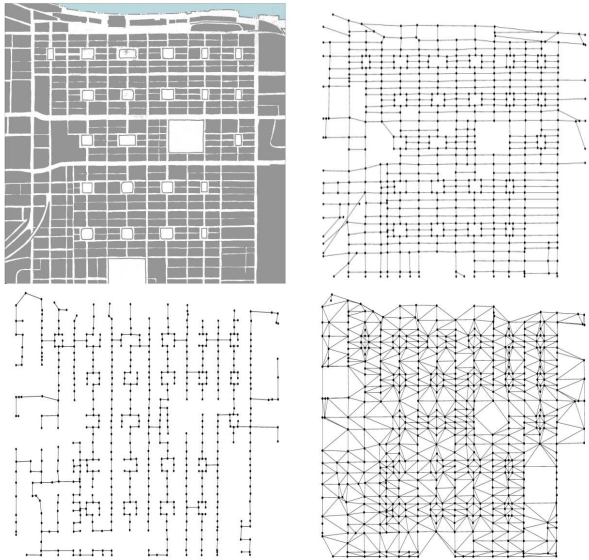}   
\caption{The urban pattern of Savannah as it appears in the 
original map (top-left), and reduced into a spatial graph 
(top-right). We also report the corresponding MST 
(bottom-left) and GT (bottom-right).}
\label{maps} 
\end{figure}

For each of the twenty cities we have constructed the respective 
MST and GT. These two bounds make also sense as regards as the 
possible evolution of a city: the most primitive forms 
are close to trees, while more complex forms involve the presence 
of cycles. We can then compare the structural properties of the 
orginal graphs representing the city with those of the two 
limiting  cases represented by MST and GT. As an example in 
Fig.~\ref{maps} in the bottom-left and in the bottom-right 
panel we report respectively the MST and the GT obtained 
for the city of Savannah. 
%

\subsection{Graph local properties} 

\begin{figure}[htb]
\includegraphics[width=8.cm]{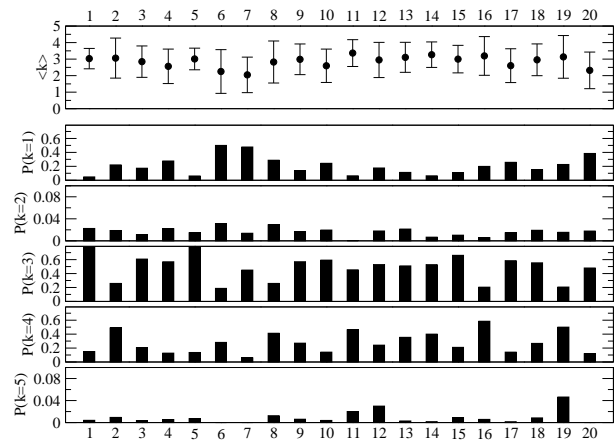}   
\caption{Average degree $\langle k \rangle $ 
and probability of having nodes with degree respectively equal to 
1, 2, 3, 4, and 5 for the twenty city considered. The cities are labeled from 1 to 20 
as reported in Table \ref{table1}. The degree distribution $P(k)$ is defined as 
$P(k) = N(k)/N$, where $N(k)$ is the number of nodes having degree $k$.}
\label{pk} 
\end{figure}

The degree of a node is the number of its direct connections to other
nodes. In terms of the adjacency matrix, the degree $k_i$ of node $i$ 
is defined as $k_i = \sum_{ j=1,N}  a_{ij}$. 
In many real networks, the degree distribution $P(k)$, defined
as the probability that a node chosen uniformly at random has
degree $k$ or, equivalently, as the fraction of nodes in the graph
having degree $k$, significantly deviates from the Poisson
distribution expected for a random graph and exhibits a power law 
(scale-free) tail with an exponent $\gamma$ taking a value between 
2 and 3 \cite{bareport,newmanreview,report}. As already mentioned 
in the introduction, we do not expect to find scale-free degree 
distributions in planar networks because the node degree is limited 
by the spatial embedding. 
In particular, in the networks of urban street patterns considered, it 
is very unprobable to find an intersection with more than 5 or 6 streets. 
In Fig.~\ref{pk} we report the average degree $\langle k \rangle $, 
%
and the degree distribution $P(k)$ for $k$ going from 1 to 5. 
The cities are labeled with an index going from 1 to 20, the 
same index we have used in Table \ref{table1}.  
In all the samples considered, the largest number of nodes 
have a degree equal to 3 or 4. Self-organized cities as Ahmedabad, Bologna,  
Cairo and Venice have $P(k=3) > P(k=4)$, 
while the inverse is true for most of the 
single-planned cities as New York, San Francisco and 
Washington, because of their square-grid structure. 
It is not the aim of this article to discuss the meaning of such differences 
in terms of their possible impacts on crucial factors for urban life, such 
as pedestrian movement, wayfinding, land-uses or other cognitive or 
behavioural matters. However, it is worth noting that, for instance, 
3-arms and 4-arms street junctions are expected to perform 
very differently in human orienteering within an urban complex system 
due to the differences in the angle widths involved in each turn 
\cite{daltonr,hillier}. 
It is also interesting to notice the significative frequency of nodes 
with degree 1 in cities as Irvine and Walnut Creek. 
Such nodes correspond to the dead-end cul-de-sac streets typical of 
the suburban early Sixties ``lollipop'' layouts, which in turn 
leads to highly debated topics in the current discussion about safety 
and liveability of modern street patterns as opposite to more traditional 
ones \cite{sou,mar}. 

Many complex networks show the presence of a large 
number of short cycles or specific motifs \cite{bareport,newmanreview,report}. 
For instance, the so called local clustering, also known as transitivity, 
is a typical property of acquaintance networks, where two individuals 
with a common friend are likely to know each other \cite{wasserman94}. 
The degree of clustering is usually quantified by the calculation of the 
{\em clustering coefficient} $C$, introduced in Ref.~\cite{ws98}, that  
is a measure of the fraction of triangles present in the network. 
Such a quantity is not suited to characterize the local properties 
of a planar graph, since by a simple counting of the number of triangles 
present in the graph it is not possible to discriminate between different 
topologies. For instance, there are cases as diverse as trees, 
square-meshes and honey-comb meshes, all having the same clustering 
coefficient equal to zero. 
Buhl et al.~have proposed a more general measure of the structure 
of cycles (not restricted to cycles of length 3) in planar graphs, 
the so called {\em meshedness coefficient} $M$ \cite{ants04}.  
The meshedness coefficient is defined as  $M = F / F_{max}$, 
where $F$ is the number of faces (excluding the external ones)  
associated with a planar graph with $N$ nodes and $K$ edges, 
and expressed by the Euler formula 
in terms of number of nodes and edges as: $F= K - N + 1$. 
$F_{max}$ is the maximum possible number of faces that is 
obtained in the maximally connected planar graph i.e. in a graph 
with $N$ nodes and $K_{max}=3N -6$ edges. Thus $F_{max}=2N-5$ and 
the meshedness coefficient can vary from zero (tree structure) 
to one (maximally connected planar graph, as in the GT \cite{note}).  

Here, we have evaluated the meshedness coefficient $M$ for each of the 
twenty cities. In addition, we have counted the cycles of 
length three, four and five by using the properties of powers of 
the adjacency matrix $A$. 
E.g., the number of cycles of length three is simply equal to 
$1/6 {\text Tr} ({ A}^3)$ ~\cite{cycles}. 
We have denoted by $C_k$ the number of cycles of length $k$ in a given 
city, and by $C^{GT}_k$ the same number in the corresponding GT.
The results are reported in Table \ref{table2}.
%
\begin{table}
\begin{tabular}{l|l|c|c|c|c|}
     & CITY           & M      & $C_3/C^{GT}_3$ & $C_4/C^{GT}_4$ & $C_5/C^{GT}_5$ \\
\tableline
1    & Ahmedabad      & 0.262  & 0.023 & 0.042 & 0.020\\
2    & Barcelona      & 0.275  & 0.019 & 0.101 & 0.019\\
3    & Bologna        & 0.214  & 0.015 & 0.048 & 0.013 \\
4    & Brasilia       & 0.147  & 0.029 & 0.027 & 0.012\\
5    & Cairo          & 0.253  & 0.020 & 0.043 & 0.019\\
6    & Irvine 1       & 0.085  & 0.035 & 0.022 & 0.005\\
7    & Irvine 2       & 0.014  & 0.007 & 0.004 & 0.001\\
8    & Los Angeles    & 0.211  & 0.002 & 0.075 & 0.011\\ 
9    & London         & 0.249  & 0.011 & 0.060 & 0.020\\
10   & New Delhi      & 0.154  & 0.011 & 0.020 & 0.011\\
11   & New York       & 0.348  & 0.024 & 0.136 & 0.028\\
12   & Paris          & 0.241  & 0.028 & 0.063 & 0.016\\
13   & Richmond       & 0.279  & 0.034 & 0.068 & 0.022\\
14   & Savannah       & 0.322  & 0.002 & 0.111 & 0.026\\   
15   & Seoul          & 0.253  & 0.021 & 0.051 & 0.021\\
16   & San Francisco  & 0.309  & 0.003 & 0.148 & 0.003\\
17   & Venice         & 0.152  & 0.016 & 0.030 & 0.010\\
18   & Vienna         & 0.242  & 0.007 & 0.063 & 0.018\\
19   & Washington     & 0.293  & 0.026 & 0.132 & 0.022\\
20   & Walnut Creek   & 0.084  & 0.000 & 0.011 & 0.003\\
\end{tabular}
\caption{Local properties of the graphs of urban street patterns. 
We report the meshedness coefficient $M$ \cite{ants04}, and 
the number $C_k$ of cycles of length $k=3,4,5$ normalized to the 
number of cycles in the GT, $C^{GT}_k$. } 
\label{table2} 
\end{table}
Three are the cities with a value of meshedness larger than 0.3: 
New York, Savannah and San Francisco. 
These represents the most complex forms of cities. On the other 
hand, Irvine and Walnut Creek with a value of $M$ lower than 0.1, 
have a tree-like structure. 
Notice that both the first and the second group of cities are examples 
of planned urban fabrics. 
On the other hand, organic patterns such as Ahmedabad, 
Cairo and Seoul also exhibit high values of mashedness, which means a considerable 
potential of local clustering. Thus, beside 
the suburban ``lollipop'' layout, both grid planned and organic 
self-organized patterns do show good local performances in terms of the local 
structural properties of the network: this is even more 
interesting if coupled with our previous finding that such two classes of 
patterns perform radically different in terms of how centrality flows over 
the network, the former exhibiting power-laws distributions while the latter 
single-scale exponential distributions \cite{porta_centrality}.
In most of the samples we have found a rather small value 
of $C_3/C^{GT}_3$ (as compared, for instance, to $C_4/C^{GT}_4$),  
denoting that triangles are not common in 
urban city patterns. This is another proof that the clustering 
coefficient $C$ alone is not a good measure to characterize the local 
properties of such networks. 
Walnut Creek, Los Angeles and Savannah are the city with the 
smallest value of  $C_3/C^{GT}_3$, while Irvine1, Richmond, Brasilia 
and Paris are the cities with the largest value of  $C_3/C^{GT}_3$. 
In 17 samples out of 20 we have found $C_4/C^{GT}_4 > C_3/C^{GT}_3$: 
Brasilia, Irvine 1 and Irvine 2 are the only cities having a prevalence 
of triangles with respect to squares. 
San Francisco, New York, Washinghton, Savannah and Barcelona 
are the cities with the largest value of $C_4/C^{GT}_4$ 
(larger than 0.1).  
Finally, concerning  $C_5/C^{GT}_5$, we have found three classes of cities. 
Samples such as Ahmedabad, Cairo, Seul and Venice having 
$C_3/C^{GT}_3 \simeq  C_5/C^{GT}_5$. Samples such as Brasilia, Irvine and  
Paris with $C_3/C^{GT}_3 > C_5/C^{GT}_5$,  
and samples as Los Angeles, Savannah and Vienna with 
$C_3/C^{GT}_3 < C_5/C^{GT}_5$.

\subsection{Graph global properties} 

\begin{figure}[htb]
\includegraphics[width=8.cm]{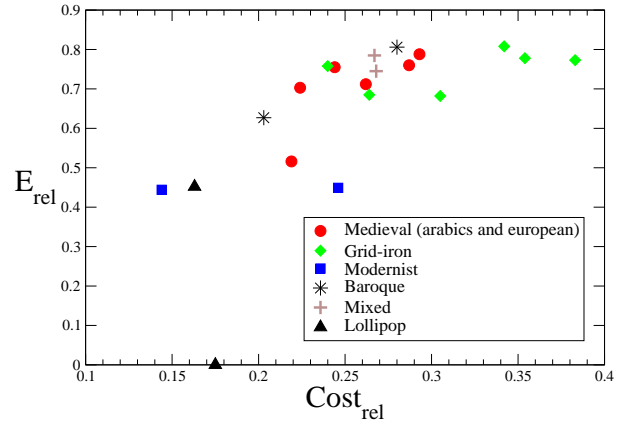}   
\caption{Relation beetween relative cost, $C_{rel}$, 
and relative efficiency, $E_{rel}$. Each point in the plot 
represent a city. 
The point of coordinates (0,0) would correspond to the 
cost/efficiency of the MST while the point (1,1) would correspond 
to the GT network. Irvine 2, having coordinates (0.175,-0.398), i.e.  
a negative value of relative efficiency, has been plotted instead 
as having coordinates (0.175,0).  
}
\label{Erel} 
\end{figure}

One of the possible mechanisms ruling the growth of an urban systems 
is the achievement of efficient pedestrian and vehicular movements on a global scale. 
This has important consequences on a number of relevant factors affecting the economic, 
environmental and social performances of cities, ranging from  accessibility to 
micro-criminality and land-uses \cite{penn}. 
The global efficiency of an urban pattern in exchanging goods, people and ideas 
should be considered a reference when the capacity of that city to support 
its internal relational potential is questioned. 
It is especially important to develop  a measure that allows the comparison between 
cases of different form and size, which poses a problem of normalization 
\cite{tek}. The global structural properties of a graph can be evaluated 
by the analysis of the shortest paths between all pairs of nodes. 
In a relational (unweighted) network the shortest path 
length between two nodes $i$ and $j$ is the 
minimum number of edges to traverse to go from $i$ to $j$.  
In a spatial (weigthed) graph, instead we define the 
shortest path length $d_{ij}$ as the smallest sum of the
edge lengths throughout all the possible paths in the graph from
$i$ to $j$ \cite{lm01,lm03}. 
In this way, both the topology and the geography 
of the system is taken into account.
As a measure of the efficiency in the communication between 
the nodes of a spatial graph, we use the so called {\em global efficiency} $E$, 
a measure defined in Ref.~\cite{lm01} as: 
\begin{equation} 
\label{efficiency}
  E  =  \frac{1}{N(N-1)}    \sum_{i, j, i \neq j } 
 \frac{ d^{Eucl}_{ij}} { d_{ij}  } 
\end{equation}
Here, $d^{Eucl}_{ij}$ is the distance between nodes $i$ 
and $j$ along a straight line, defined in Section \ref{section_mstgt}, 
and we have adopted a normalization 
recently proposed for geographic networks \cite{vragovic}.  
Such a normalization captures to which extent the connecting route 
between $i$ and $j$ deviates from the virtual straight line.  
%
\begin{table}
\begin{tabular}{l|l|c|c|c}
     & CITY           &  $E$  &$E^{MST}$& $E^{GT}$\\
\tableline
1    & Ahmedabad      & 0.818 &  0.351  &  0.944  \\
2    & Barcelona      & 0.814 &  0.452  &  0.930  \\
3    & Bologna        & 0.799 &  0.473  &  0.936  \\
4    & Brasilia       & 0.695 &  0.503  &  0.931  \\
5    & Cairo          & 0.809 &  0.385  &  0.943  \\
6    & Irvine 1       & 0.755 &  0.604  &  0.943  \\
7    & Irvine 2       & 0.374 &  0.533  &  0.932  \\
8    & Los Angeles    & 0.782 &  0.460  &  0.930  \\
9    & London         & 0.803 &  0.475  &  0.936  \\
10   & New Delhi      & 0.766 &  0.490  &  0.930  \\
11   & New York       & 0.835 &  0.433  &  0.931  \\
12   & Paris          & 0.838 &  0.473  &  0.938  \\
13   & Richmond       & 0.800 &  0.502  &  0.939  \\
14   & Savannah       & 0.793 &  0.341  &  0.922  \\
15   & Seoul          & 0.814 &  0.444  &  0.941  \\
16   & San Francisco  & 0.792 &  0.448  &  0.893  \\
17   & Venice         & 0.673 &  0.386  &  0.943  \\
18   & Vienna         & 0.811 &  0.423  &  0.937  \\
19   & Washington     & 0.837 &  0.452  &  0.930  \\
20   & Walnut Creek   & 0.688 &  0.481  &  0.938  \\
\end{tabular}
\caption{The efficiency $E$ of each city is compared 
to the minimum and maximum values of the efficiency obtained 
respectively for the MST and the GT.   
The cities are labeled from 1 to 20 as in Table \ref{table1}
} 
\label{table3} 
\end{table}
In Table \ref{table3} we report the values of efficiency obtained for each 
city and for the respective MST and the GT. The values of $E^{MST}$ and 
$E^{GT}$ serve to normalize the results, being respectively 
the minimum and the maximum value of efficiency that can be obtained 
in a planar graph having the same number of nodes as in the original graph of 
the city. 
Notice that Irvine 2 is the only case in which $E < E^{MST}$. This is 
simply due to the fact that Irvine 2 is the only city whose corresponding 
graph is not connected. 
Consequently, the MST has a smaller number of edges 
but a higher value of efficiency because it is, by definition, 
a connected graph. 
The main result is that the cities considered, despite their inherent
differences, achieve a relatively high value of efficiency, 
which is in most of the cases about $80\%$ of the maximum 
value of the efficiency in a planar graph, $E^{GT}$. 
Following Ref.~\cite{ants04} we define the relative efficiency 
$E_{rel}$ as: 
\begin{equation} 
\label{efficiencyrelative}
  E_{rel} =  \frac{ E - E^{MST} } {E^{GT} - E^{MST}} 
\end{equation} 
Of course, the counterpart of an increase in efficiency is 
an increase in the cost of construction, i.e. an increase 
in the number and length of the edges. The cost of construction 
can be quantified by using the measure $Cost$ defined in 
formula (\ref{eq:cost}).    
Given a set of $N$ nodes, the shortest (minimal cost) 
planar graph that connects all nodes correspons to the MST, 
while a good approximation for the maximum cost planar graph 
is given by the GT. 
We thus define a normalized cost measure, $Cost_{rel}$, as: 
\begin{equation} 
\label{efficiencyrelative}
  Cost_{rel} =  \frac{ Cost - Cost^{MST} } {Cost^{GT} - Cost^{MST}} 
\end{equation}
By definition the MST has a relative cost $Cost_{rel}=0$, while 
GT has $Cost_{rel}=1$. 
In Fig.~\ref{Erel} we plot for each city $E_{rel}$ as a function 
of $Cost_{rel}$. 
%
The cities can be a-priori divided into different classes: 
1) medieval fabrics, including both arabic (Ahmedabad and Cairo) 
and european (Bologna, London, Venice, and Vienna); 
2) grid-iron fabrics (Barcelona, Los Angeles, New York, 
Richmond, Savannah and San Francisco); 
3) modernist fabrics (Brasilia and Irvine 1); 
4) baroque fabrics (New Delhi and Washington); 
5) mixed fabrics (Paris and Seoul); 
6) ``lollipop'' layouts (Irvine 2 and Walnut Creek). 
We shall see that the plot $E_{rel}$ vs. $C_{rel}$ has 
a certain capacity to characterize the different classes 
of cities listed above. 
The plot indicates an overall increasing behavior of $E_{rel}$ 
as function of $Cost_{rel}$, with a saturation at 
$E_{rel} \sim 0.8$ for values of $Cost_{rel} > 0.3$. 
Grid-iron patterns exhibit a high value of 
relative efficiency, about $70-80\%$ of the efficiency of the GT, 
with a relative cost which goes from $0.24$ to $0.4$. 
The three grid-iron cities (New York, Savannah and San Francisco)
with the largest value of efficiency, $E_{rel} \sim 0.8$, have 
respectively a cost equal to 0.342, 0.354 and 0.383. 
Medieval patterns have in general a lower cost and efficiency 
than grid-iron patterns althouh, in some cases as Ahmedabad and Cairo 
(the two medieval cities with the largest efficiency), 
they can also reach a value of $E_{rel} \sim 0.8$ with 
a smaller cost equal to 0.29. 
Modernist and ``lollipop'' layouts are those with the smallest 
value of Cost but also with the smallest value of efficiency. 

\section{Conclusions} 

We have proposed a method to characterize both the local and the 
global properties of spatial graphs representing urban street 
patterns. 
Our results show that a comparative analysis on the structure of 
different world cities is possible by the introduction of two limiting 
auxiliary graphs, the MST and the GT,  
A certain level of structural similarities across cities as well as 
some difference are well captured by counting cycles and by  
measuring normalized efficiency and cost of the graphs.  
The method can be applied to other planar graphs of different 
nature, as highway or railway networks.

{\bf Acknowledgment.} 
\noindent
We thank J. Buhl, P. Crucitti, R.V. Sol\'e and S. Valverde, 
for many helpful discussions and suggestions.


\begin{thebibliography}{99}
\bibitem{note} 
In the computed GT, the graphs we have obtained have less than $3N -6$ edges. 
This is due to the fact that we have considered all the edges as 
being straight-lines connecting the two end-nodes. For such a reason some of 
edges connecting the external nodes (the nodes on the border of the unit square) 
cannot be placed without causing edge crossings. However, the number of these 
edges is of the order of $\sqrt{N}$, so that in large GT graphs the 
meshedness is close to 1.
\end{thebibliography}
\end{document}